\documentclass[aps,prl,reprint,superscriptaddress,showpacs,longbibliography,footnoteinbib]{revtex4-1}
\usepackage[latin9]{inputenc}
\usepackage{geometry}
\usepackage{color}
\usepackage{bm}
\usepackage{amsmath}
\usepackage{amssymb}
\usepackage{graphicx}
\usepackage[unicode=true,
 bookmarks=false,
 breaklinks=false,pdfborder={0 0 1},backref=false,colorlinks=true]
 {hyperref}
\hypersetup{pdftitle={Title},
 plainpages=false,pdfpagelabels,linkcolor=blue,urlcolor=blue,citecolor=blue,pdfdisplaydoctitle=true,pdfduplex=DuplexFlipLongEdge}

\makeatletter

\newcommand{\lyxdot}{.}

\usepackage{units}\usepackage{wasysym}

\definecolor{orange}{rgb}{0.50, 0.20, 0.0}

\usepackage{bm}
\usepackage{braket}

\makeatother

\begin{document}
\title{Entanglement entropy scaling in critical phases of 1D quasiperiodic
systems}
\author{Miguel Gonçalves}
\affiliation{CeFEMA-LaPMET, Departamento de Física, Instituto Superior Técnico,
Universidade de Lisboa, Av. Rovisco Pais, 1049-001 Lisboa, Portugal}
\affiliation{Centro de Física das Universidades do Minho e Porto, Departamento
de Física e Astronomia, Faculdade de Ciências, Universidade do Porto,
4169-007 Porto, Portugal}
\begin{abstract}
We study the scaling of the entanglement entropy in different classes
of one-dimensional fermionic quasiperiodic systems with and without
pairing, focusing on multifractal critical points/phases. We find
that the entanglement entropy scales logarithmically with the subsystem
size $N_{A}$ with a proportionality coefficient $\mathcal{C}$, as
in homogeneous critical points, apart from possible additional small
oscillations. In the absence of pairing, we find that the entanglement
entropy coefficient $\mathcal{C}$ is non-universal and depends significantly
and non-trivially both on the model parameters and electron filling,
in multifractal critical points. In some of these points, $\mathcal{C}$
can take values close to the homogeneous (or ballistic) system, although
it typically takes smaller values. We find a close relation between
the behaviour of the entanglement entropy and the small-$q$ (long-wavelength)
dependence of the momentum structure factor $\mathcal{S}(q)$. $\mathcal{S}(q)$
increases linearly with $q$ as in the homogeneous case, with a slope
that grows with $\mathcal{C}$. In the presence of pairing, we find
that even the addition of small anomalous terms affects very significantly
the scaling of the entanglement entropy compared to the unpaired case.
In particular, we focused on topological phase transitions for which
the gap closes with either extended or critical multifractal states.
In the former case, the scaling of the entanglement entropy mirrors
the behaviour observed at the critical points of the homogeneous Kitaev
chain, while in the latter, it shows only slight deviations arising
at small length scales. In contrast with the unpaired case, we always
observe $\mathcal{C}\approx1/6$ for different critical points, the
known value for the homogeneous Kitaev chain with periodic boundary
conditions.

\end{abstract}
\maketitle
\noindent\begin{minipage}[t]{1\columnwidth}%
\global\long\def\ket#1{\left| #1\right\rangle }%

\global\long\def\bra#1{\left\langle #1 \right|}%

\global\long\def\kket#1{\left\Vert #1\right\rangle }%

\global\long\def\bbra#1{\left\langle #1\right\Vert }%

\global\long\def\braket#1#2{\left\langle #1\right. \left| #2 \right\rangle }%

\global\long\def\bbrakket#1#2{\left\langle #1\right. \left\Vert #2\right\rangle }%

\global\long\def\av#1{\left\langle #1 \right\rangle }%

\global\long\def\tr{\text{tr}}%

\global\long\def\Tr{\text{Tr}}%

\global\long\def\pd{\partial}%

\global\long\def\im{\text{Im}}%

\global\long\def\re{\text{Re}}%

\global\long\def\sgn{\text{sgn}}%

\global\long\def\Det{\text{Det}}%

\global\long\def\abs#1{\left|#1\right|}%

\global\long\def\up{\uparrow}%

\global\long\def\down{\downarrow}%

\global\long\def\vc#1{\mathbf{#1}}%

\global\long\def\bs#1{\boldsymbol{#1}}%

\global\long\def\t#1{\text{#1}}%
\end{minipage}

\section{Introduction}

The study of entanglement in many-body systems has been a great example
of success where remarkable new insights were attained by applying
concepts from quantum information theory to condensed matter systems.
Among them, the universal scaling of the entanglement entropy in conformal
critical points of one-dimensional systems stands out \citep{PhysRevLett.90.227902,Calabrese_2004,Calabrese_2009,RevModPhys.80.517}.
At these critical 1D systems, the entanglement entropy of a finite
subsystem of an infinite system with $N_{A}$ sites, scales as $S=\mathcal{C}\log(N_{A})+{\rm cte}$,
where $\mathcal{C}=c/3$ for systems with periodic boundary conditions
(we refer below to systems with periodic boundaries unless otherwise
stated) and $c$ is the universal central charge of the corresponding
conformal field theory. For a free fermion chain (that can be mapped
to the XX model, through a Jordan-Wigner transformation), $c=1$.
On the other hand, at the critical points of the topological phase
transitions in the Kitaev chain (or XY model), we have $c=1/2$. In
the presence of a gap, the entanglement entropy also scales logarithmically
up to a length scale of the order of the inverse energy gap, above
which it saturates. Given the universality of the entanglement entropy
scaling, it does not depend on model details for a group of models
that are all described by the same low-energy conformal field theory.
This includes the addition of inhomogeneities such as disorder can
therefore still lead to universal behaviour \citep{PhysRevLett.93.260602,PhysRevA.75.052329,Refael_2009,PhysRevB.89.115104,PhysRevB.90.104204,PhysRevB.102.214201}. 

Other very interesting inhomogeneous systems for which entanglement
has been less studied are quasiperiodic systems. These systems offer
a wide range of exciting physics, from interesting localization properties
\citep{AubryAndre,Roati2008,Lahini2009,Schreiber842,Luschen2018,Huang2016a,PhysRevLett.120.207604,Park2018,PhysRevB.100.144202,Fu2020,Wang2020,goncalves2020}
to topological features and edge physics \citep{Kraus2012,PhysRevLett.109.116404,Verbin2013}.
Interest in quasiperiodic systems has been recently renewed due to
progress in experiments in optical lattices \citep{PhysRevA.75.063404,Roati2008,Modugno_2009,Schreiber842,Luschen2018}
and metamaterials \citep{Lahini2009,Kraus2012,Verbin2013,PhysRevB.91.064201,Wang2020},
and the increased focus on moiré systems \citep{Balents2020,Andrei2021}. 

In one dimension, quasiperiodic systems host rich localization physics,
transitions between extended ballistic phases with plane-wave-like
eigenstates and localized phases, where the wave function is exponentially
localized in real-space \citep{AubryAndre,Roati2008,Lahini2009,Schreiber842,Luschen2018}.
At the critical point, the eigenstates are multifractal both in real
and momentum space. For some systems, non-fine-tuned phases with multifractal
eigenstates can arise \citep{PhysRevLett.110.146404,Liu2015,PhysRevB.93.104504,PhysRevLett.123.025301,PhysRevLett.125.073204,CadeZ2019,anomScipost,goncalves2023critical}.
The entanglement entropy has been previously studied in quasiperiodic
systems \citep{PhysRevB.93.184204,PhysRevB.100.195143,PhysRevA.87.043635,PhysRevB.102.064204,PhysRevB.103.024202,PhysRevB.103.075124}.
For the half-filled paradigmatic Aubry-André model \citep{AubryAndre},
it was found to scale logarithmically with subsystem size at the extended
phase and critical point, and to saturate in the localized phase at
length scales larger than the localization length \citep{PhysRevA.87.043635,PhysRevB.97.125116,PhysRevB.102.064204}.
In the extended phase, the entanglement entropy was found to behave
as in the homogeneous case, scaling with $\mathcal{C}\approx1/3$.
At the critical point, however, a different coefficient $\mathcal{C}\approx0.26$
was observed \citep{PhysRevB.102.064204}. The entanglement entropy
of a generalized Aubry-André model with long-range (power-law decaying)
hoppings was also studied in Ref.$\,$\citep{PhysRevB.103.075124},
at half-filling. For some parameters, it was found that $\mathcal{C}\neq1/3$
even in the extended phase. In critical phases, $\mathcal{C}$ was
always found to be significantly lower. This raises the question on
whether extended and critical phases can always be distinguished based
on the scaling of the entanglement entropy. Moreover, a lot remains
to explore on the behaviour of the entanglement entropy in more generic
regimes, including different fillings and different critical points
and critical phases. In the presence of pairing, the entanglement
entropy has not been studied so far on critical points, to our knowledge.

In this work, using numerically exact methods, we study the behaviour
of the entanglement entropy in different classes of one-dimensional
quasiperiodic systems both with and without pairing terms. We focus
on multifractal critical points/phases and find that even though the
entanglement entropy scales logarithmically with the subsystem size,
apart from possible small oscillations, the coefficients $\mathcal{C}$
depend significantly on the model parameters and electron filling
in the absence of pairing. In fact, we observe that $\mathcal{C}\approx1/3$
at some critical points, with the entanglement entropy behaving similarly
to the homogeneous case. Interestingly, the scaling of the entanglement
entropy is closely related with the long-wavelength (low-energy) behaviour
of the momentum structure factor $\mathcal{S}(q)$, growing linearly
with $q$, with a slope that increases with $\mathcal{C}$. 

To understand the impact of the pairing terms, we studied the quasiperiodic
Kitaev chain, focusing on critical points of topological phase transitions.
In this case, we found that the entanglement entropy behaves very
similarly to the homogeneous Kitaev chain, where $\mathcal{C}=1/6$,
irrespectively of the model parameters. This is true no matter whether
extended or multifractal states are present at the critical point,
with the most significant deviations from the $\mathcal{C}=1/6$ behaviour
only occurring for the latter case, at small length scales.

Our results show that while in the absence of pairing the scaling
of the entanglement entropy can vary quite significantly in critical
phases and even become very similar to the homogeneous case, the addition
of pairing terms, even if small, is highly relevant and the scaling
tends to become very similar to the homogeneous Kitaev critical chain
at sufficiently large length scales.

\section{Model and methods}

\begin{figure}[h]
\centering{}\includegraphics[width=1\columnwidth]{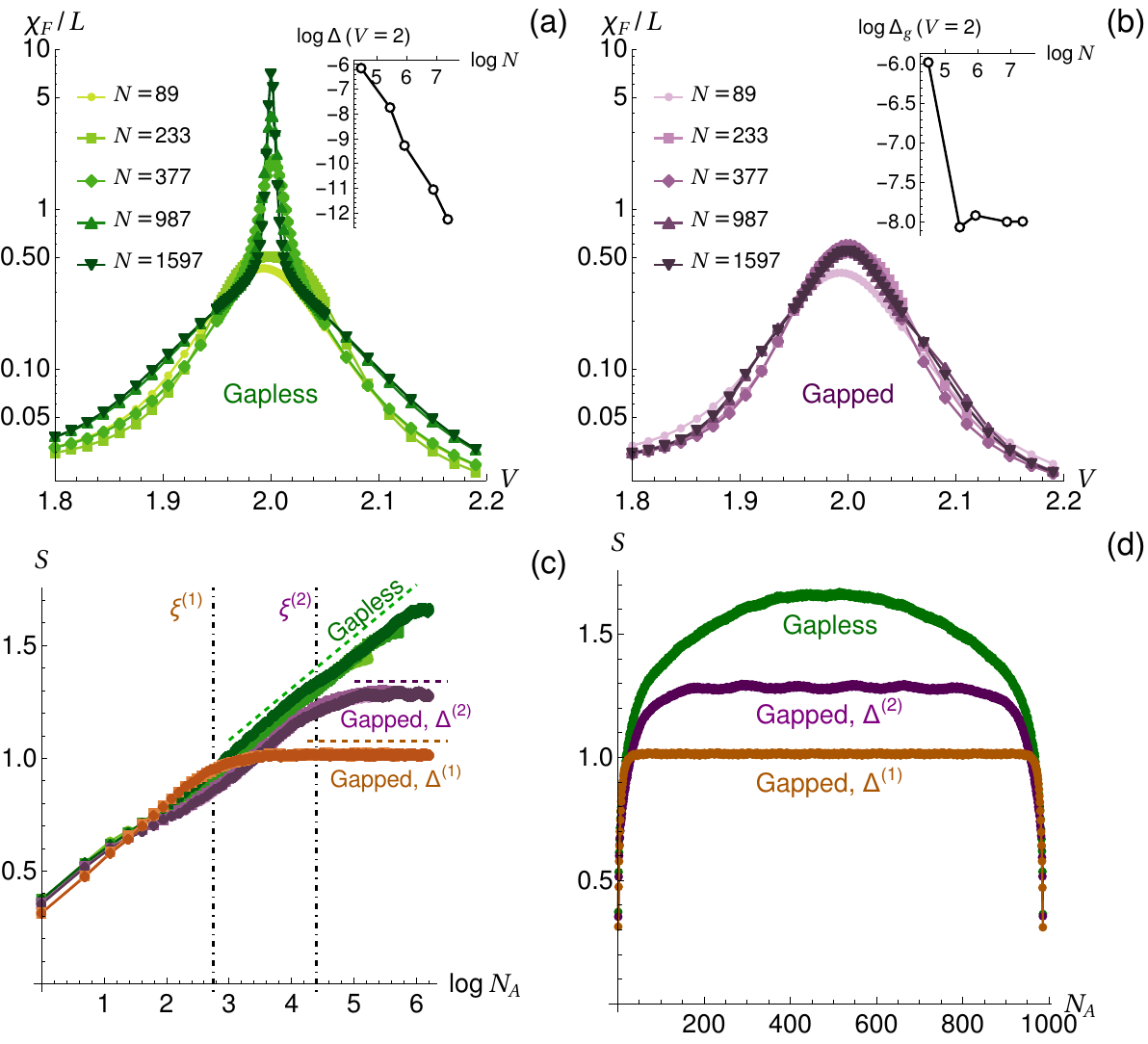}\caption{Results for the Aubry-André model. (a,b) $\chi_{F}/L$ at fillings
$\rho=1/4$ (a) and $\rho=\mod(57\tau,1)$ (b), for fixed twist $\kappa=1.7$
and $\phi=1.123/N$. The insets show the energy gap as a function
of system size for $V=2$. (c) Entanglement entropy as function of
$\log N_{A}$ up to $N_{A}=N/2$ for $V=2$, for $\rho=1/4$ (green),
$\rho^{(1)}=\mod(10\tau,1)$ (orange) and $\rho^{(2)}=\mod(57\tau,1)$
(purple), and for system sizes $N=\{377,610,987\}$ (respectively
from lighter to darker color shade). Averages over $N_{c}=1800$ configurations
were made. (d) Entanglement entropy for $N_{A}\in[1,N]$, for $N=987$,
for the same parameters as in (c). \label{fig:1}}
\end{figure}

\textit{Models.---} We consider the following class of Hamiltonians, 

\begin{equation}
\begin{aligned}\mathcal{H}= & -t\sum_{n}(c_{n}^{\dagger}c_{n+1}+c_{n+1}^{\dagger}c_{n})+\sum_{n}V_{n}c_{n}^{\dagger}c_{n}\\
 & +\Delta\sum_{n}(c_{n}c_{n+1}+c_{n+1}^{\dagger}c_{n}^{\dagger})
\end{aligned}
\label{eq:Hamiltonian}
\end{equation}

\noindent where $c_{n}^{\dagger}$ creates an electron at site $n$
and $V_{n}$ will be considered to be either of the following quasiperiodic
potentials: $V_{n}^{(1)}=V\cos(2\pi\tau n+\phi)-\mu$ and $V_{n}^{(2)}=\cos(2\pi\tau n+\phi)/[1+\alpha\cos(2\pi\tau n+\phi)]-\mu$.
In the definitions of $V_{n}$, $\tau$ is an irrational number that
we take to be $\tau=(\sqrt{5}-1)/2$, $\phi$ is a (phase) shift in
the quasiperiodic potential and $\mu$ is the chemical potential.
In what follows, energy will be measured in units of $t$. We will
study three particular cases of the Hamiltonian in Eq.$\,$\ref{eq:Hamiltonian}:
the Aubry-André model ($\Delta=0$ and $V_{n}^{(1)}$) \citep{AubryAndre},
the Ganeshan-Pixley-Sarma (GPS) model, with $\Delta=0$ and $V_{n}^{(2)}$
\citep{PhysRevLett.114.146601}, and the quasiperiodic Kitaev chain
model, $\Delta\neq0$ with $V_{n}^{(1)}$. For the GPS model, the
localization phase diagram is analytically known, hosting extended,
localized and critical phases \citep{PhysRevLett.114.146601,AnomMobEdge}.
For the quasiperiodic Kitaev chain, the localization and topological
phase diagrams in the $(V,\Delta)$ plane have also been explored
\citep{PhysRevLett.110.176403,DeGottardi_2011,PhysRevLett.110.146404,PhysRevB.93.104504}.
In particular, for $\mu=0$, a topological phase transition between
a topological phase with zero-energy majorana modes and a trivial
phase takes place. At this critical point there is also a localization
transition between phases with critical multifractal eigenstates (topological)
and localized eigenstates (trivial). 

In order to simulate finite systems we followed the usual procedure
carried out for QPS \citep{PhysRevLett.43.1954,Szabo2018,Wang2020a}:
for each size $N=F_{n}$, we take $\tau$ to be a rational approximant
of the inverse of the golden ratio, $\tau=\tau_{n}=F_{n-1}/F_{n}$,
where $F_{n}$ is the $n$-th Fibonacci number. Note that in this
way the finite system is incommensurate (the system's unit cell for
$\tau=\tau_{n}$ has precisely $N$ sites, the total system size)
and we may apply periodic boundary conditions (PBC) without defects. 

\textit{Entanglement entropy and fidelity.---} We can write the quadratic
fermionic Hamiltonian in Eq.$\,$\ref{eq:Hamiltonian} in the Nambu
representation,

\begin{equation}
\mathcal{H}=\frac{1}{2}\bm{C}^{\dagger}\bm{H}\bm{C},\textrm{ }\bm{H}=\left(\begin{array}{cc}
\bm{h} & \bm{\Delta}\\
\bm{\Delta}^{\dagger} & -\bm{h}^{T}
\end{array}\right)
\end{equation}
, where $\bm{C}=(c_{1},\cdots,c_{N},c_{1}^{\dagger},\cdots,c_{N}^{\dagger})^{T}$
is a Nambu vector and 

\begin{equation}
\bm{h}=\left(\begin{array}{ccccc}
V_{0} & -t & 0 & \cdots & -t\\
-t & V_{1} & -t & \cdots & 0\\
0 & -t & \ddots & \cdots & \vdots\\
\vdots & \vdots & \cdots & V_{N-2} & -t\\
-t & 0 & \cdots & -t & V_{N-1}
\end{array}\right)
\end{equation}

\begin{equation}
\bm{\Delta}=\left(\begin{array}{ccccc}
0 & \Delta & 0 & \cdots & -\Delta\\
-\Delta & 0 & \Delta & \cdots & 0\\
0 & -\Delta & \ddots & \cdots & \vdots\\
\vdots & \vdots & \cdots & 0 & \Delta\\
\Delta & 0 & \cdots & -\Delta & 0
\end{array}\right)
\end{equation}
with these matrices satisfying $\bm{h}=\bm{h}^{\dagger}$ and $\bm{\Delta}=-\bm{\Delta}^{T}$.
The single-particle correlation matrix can then be defined as $\bm{\chi}=\tr\Big(Z^{-1}e^{-\beta\mathcal{H}}\bm{C}\bm{C}^{\dagger}\Big)=\bm{\mathbb{I}}-n_{F}(\bm{H})$,
where $\beta$ is the inverse temperature, $Z$ is the partition function
and $n_{F}$ is the Fermi-Dirac distribution. In order to compute
the entanglement entropy, we consider our subsytem $A$ to contain
the first $N_{A}$ sites of the full chain containing $N$ sites.
The entanglement entropy can then be computed as 

\begin{equation}
S_{A}=-\Tr[\bm{\chi}_{A}\ln\bm{\chi}_{A}],\textrm{ }\bm{\chi}_{A}=[\bm{\chi}]_{i,j\in\mathcal{A}\cup(\mathcal{A}+N)},
\end{equation}

\noindent where $i,j$ are entries of matrix $\bm{\chi}$, $\mathcal{A}$
is the set of site indexes of subsytem $A$ and $\mathcal{A}+N$ are
the indices obtained by summing $N$ to each of the site indices in
$\mathcal{A}$. 

In what follows, we average the results for the entanglement entropy
over random uniformly distributed realizations of phase $\phi$, unless
otherwise stated. For the calculations in the Aubry-André and GPS
models, we also average over random twisted boundary conditions, by
closing the boundaries with a phase twist $e^{ik}$ ($k=0$ for PBC)
\citep{PhysRevLett.120.207604}. We estimate the error bars of the
average entanglement entropy through the error of the mean. 

For a 1D critical system whose continuum limit is a conformal field
theory, we have that \citep{Calabrese_2004,Calabrese_2009}

\begin{equation}
S=\mathcal{C}\log\Big(\frac{N}{\pi}\sin(\pi N_{A}/N)\Big)+C',\label{eq:calebrese}
\end{equation}
where $N_{A}$ is the number of sites in subsystem $A$, in which
case the central charge of the conformal field theory is given by
$\mathcal{C}/3$, for periodic boundary conditions. For $N_{A}\ll N$
we therefore have $S=\mathcal{C}\log N_{A}$. For the homogeneous
fermionic chain with no anomalous terms ($V=\Delta=0$), we have that
$\mathcal{C}=1/3$. For the Kitaev chain ($V=0,\Delta\neq0$), on
the other hand, we have that $\mathcal{C}=1/6$ because the central
charge takes half the value of the homogeneous chain. For $\Delta=0$
and $V\neq0$, the $\mathcal{C}=1/3$ scaling was found to hold for
$V$ within the extended phase \citep{PhysRevB.102.064204} in the
half-filled Aubry-André model. However, Ref.$\,$\citep{PhysRevB.103.075124}
found regimes where $\mathcal{C}$ was smaller even for electron fillings
within the extended phase for a long-range Aubry-André model, in the
presence of multifractal to extended mobility edges. In the localized
phase, $S$ only grows logarithmically for $N_{A}\lesssim\xi$, where
$\xi$ is the localization length \citep{PhysRevB.97.125116,PhysRevB.103.075124,PhysRevB.102.064204}.
At critical points containing multifractal eigenstates, $S$ was found
to still scale logarithmically with $N_{A}$, although with significantly
smaller $\mathcal{C}$, accompanied by possible log-periodic oscillations
\citep{goncalves2023shortrange}, as previously observed for aperiodic
quantum spin chains \citep{Juhasz_2007,Igloi_2007}.

We will also compute the ground state fidelity, defined as \citep{PhysRevE.74.031123,doi:10.1142/S0217979210056335}

\begin{equation}
F(\Psi(\lambda),\Psi(\lambda'))=|\braket{\Psi(\lambda)}{\Psi(\lambda')}|,\label{eq:fid-1}
\end{equation}

\noindent where $\ket{\Psi(\lambda)}$ is the system's ground-state
and $\lambda$ is some selected parameter of the model. In the case
of non-interacting fermions, the overlap in Eq.$\,$\ref{eq:fid-1}
can be simply computed through $F(\Psi(\lambda),\Psi(\lambda'))=|\det(\bm{\Phi}_{\lambda}^{\dagger}\bm{\Phi}_{\lambda'})|$,
where $\bm{\Phi}_{\lambda}$ is a $N\times M$ matrix containing the
$M$ filled single-particle eigenstates in its columns. Taking $\lambda'=\lambda+\delta\lambda$,
we can compute the leading-order term of the fidelity as $\delta\lambda\rightarrow0$.
This term corresponds to the fidelity susceptibility $\chi_{F}$,
which can be defined as \citep{PhysRevE.76.022101,doi:10.1142/S0217979210056335} 

\begin{equation}
\chi_{F}=-\frac{\pd^{2}F}{\pd(\delta\lambda)^{2}}=\lim_{\delta\lambda\rightarrow0}\frac{-2\log F}{(\delta\lambda)^{2}}.\label{eq:chi_F}
\end{equation}

In the following, we choose $\delta\lambda$ small enough for $\chi_{F}$
to be converged. The finite-size scaling of $\chi_{F}$ can be used
to tackle critical points and critical exponents. At a critical point,
the maximum of $\chi_{F}$, typically scales super extensively and
its maximizant approaches the critical point as the system size is
increased \citep{Fidelity_QPT}. 

\textit{Localization probes.---} To inspect the localization properties,
we compute inverse participation ratios (IPR). For each single-particle
eigenstate $\ket{\psi_{\alpha}}=\sum_{n}\psi_{n}^{\alpha}\ket n$,
where $\{\ket n\}$ is a basis localized at each site (or the Nambu
basis for $\Delta\neq0$), the generalized IPR is given by ${\rm IPR}_{\alpha}(q)=(\sum_{n}|\psi_{n}^{\alpha}|^{2})^{-q}\sum_{n}|\psi_{n}^{\alpha}|^{2q}$
\citep{Aulbach_2004}. In general we have ${\rm IPR}_{\alpha}(q)\sim N^{-\tau(q)}$,
where $\tau(q)=D_{r}(q)(q-1)$. In the extended phase, we have $D_{r}(q)=D$,
while in the localized phase the IPRs become $L$-independent for
$L$ sufficiently larger than the localization length and therefore
$D_{r}(q)=0$. Critical points/phases in 1D quasiperiodic systems
have multifractal eigenstates, in which case $D_{r}(q)$ is a non-linear
function of $q$. For the quasiperiodic Kitaev chain, we also define
a momentum space generalized IPR. In the diagonal basis defined as
$\gamma_{\alpha}=\sum_{i}u_{\alpha,i}c_{i}^{\dagger}+v_{\alpha,i}c_{i}$,
we define $\tilde{u}_{\alpha,k}=L^{-1/2}\sum_{j}e^{2\pi jk/L}u_{\alpha,i}$
and $\tilde{v}_{\alpha,k}=L^{-1/2}\sum_{j}e^{2\pi jk/L}v_{\alpha,i}$.
The momentum-space generalized IPR is then given by ${\rm IPR}_{k}(q)=\sum_{k}(|\tilde{u}_{\alpha,k}|^{2q}+|\tilde{v}_{\alpha,k}|^{2q})\sim L^{-\tau_{k}(q)}$.
In this case, we have $\tau_{k}(q)=D_{k}(q)(q-1)$, with $D_{k}(q)=0$
for ballistic extended states, $D_{k}(q)=1$ for localized states
and $D_{k}(q)$ is again a non-linear function of $q$ for multifractal
states. In what follows, we mostly compute the real-space IPR and
set $q=2$, unless otherwise stated, defining ${\rm IPR}={\rm IPR}(q=2)$. 

\begin{figure*}[t]
\centering{}\includegraphics[width=1\textwidth]{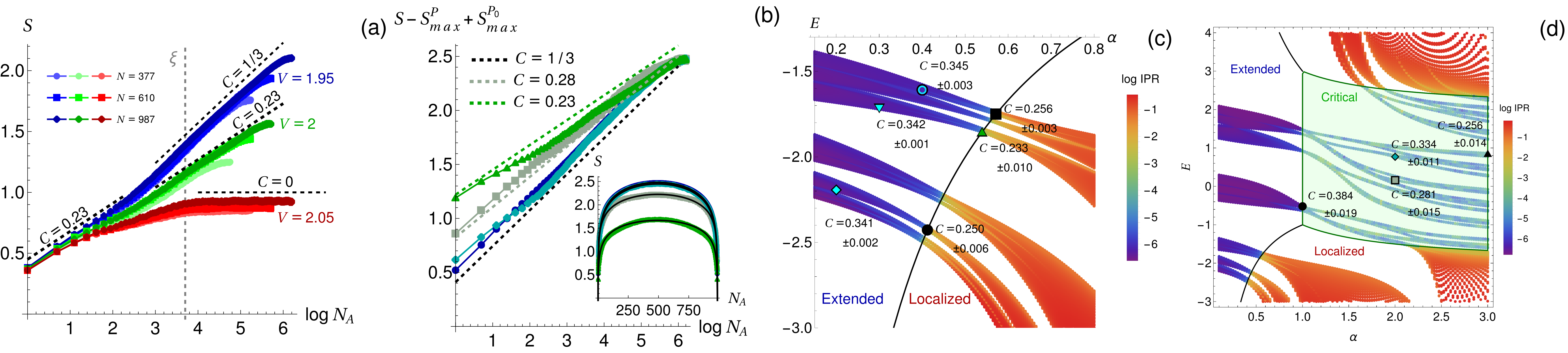}\caption{(a) Entanglement entropy $S$ for the Aubry-André model, for filling
$\rho=1/4$. The blue, green and red data points correspond to $V=1.95$
(extended), $V=2$ (critical) and $V=2.05$ (localized), respectively.
The dashed lines are guides to the eye for the scaling of the entanglement
entropy. The gray dashed line indicates $\log\xi=2.7$, where $\xi$
is approximately the exact correlation length for $V=1.95$ ($\xi=1/\log (2/V)$)
and $V=2$ ($\xi=1/\log (V/2)$). (b) $S$ for different critical points
{[}$V=1,\alpha=2,\rho=0.47$ (green); $V=1,\alpha=2,\rho=0.538$ (light
blue); $V=1,\alpha=0.539,\rho=0.25$ (gray){]} and an extended point
($\rho=0.25,\alpha=0,V=1$). $S$ was shifted so that the maxima coincide
for all the analysed points, where $P_{0}$ is the extended point
and $P$ is the point being analysed. These points are signaled in
(c,d), with the plot markers matching the ones in this figure. The
dashed lines show the slopes $\mathcal{C}$ obtained by fitting $S$
to Eq.$\,$\ref{eq:calebrese} for all $N_{A}$, where the fits are
shown in the inset. (c) ${\rm IPR}$ for the GPS model around an extended-to-localized
transition, for $L=987$ and $V=1$, along with fitting results for
$\mathcal{C}$ at different points. The black line denotes the analytical
mobility edge between the extended and localized phases. (d) ${\rm IPR}$
for the same model, in a range of parameters containing the critical
phase, together with the value of $\mathcal{C}$ extracted from the
fit within this phase. The error bars correspond to the standard deviation
of the fitted $\mathcal{C}$ for the three larger considered sizes,
$N=\{610,987,1597\}$. For all the calculations of the entanglement
entropy in (a-d), we averaged over $1800$ configurations of shits
and twists for $N=\{377,610,987\}$ and $252$ configurations for
$N=1597$. \label{fig:2}}
\end{figure*}

\textit{Topological invariant.---} Finally, in order to inspect topological
properties of the quasiperiodic Kitaev chain, we make use of the topological
invariant  introduced in Ref.$\,$\citep{DeGottardi_2011}. By writing
the Hamiltonian in terms of two species of Majorana fermions, the
equations of motion for zero energy Majorana modes can be written
in terms of transfer matrices (see Appendix \ref{sec:AppendixB} for
details). The transfer matrix for one of the species of modes is defined,
for a system of length $N$, as $\textrm{\ensuremath{\mathcal{A}_{N}}}=\prod_{n=1}^{N}A_{n}$,
where $A_{n}^{ij}=V_{n}/(\Delta+t)\delta_{i,0}\delta_{j,0}+(\Delta-t)/(\Delta+t)\delta_{i,0}\delta_{j,1}+\delta_{i,1}\delta_{j,0}$.
For the other species, the transfer matrix is $B_{n}=\sigma_{x}A_{n}^{-1}\sigma_{x}$.
A topological invariant can then be defined as in Ref.$\,$\citep{DeGottardi_2011},

\begin{equation}
\nu_{T}=-(-1)^{n_{f}},\label{eq:topo_invariant-1}
\end{equation}

\noindent where $n_{f}$ is the number of eigenvalues of $\textrm{\ensuremath{\mathcal{A}_{N}}}$
with magnitude smaller than unity. Inside a topological phase $\nu_{T}=-1(n_{f}=0,2)$,
while in a trivial phase $\nu_{T}=1(n_{f}=1)$. 

\textit{Structure factor.---} In order to compare the results for
the entanglement entropy scaling with other physical observables,
we also compute the momentum structure factor, defined as

\begin{equation}
\mathcal{S}(q)=N^{-1}\sum_{j,l}[\langle n_{j}n_{l}\rangle-\langle n_{j}\rangle\langle n_{l}\rangle]e^{\textrm{i}q(j-l)}\label{eq:Sq}
\end{equation}

In the non-interacting limit that we study in this manuscript, $\mathcal{S}(q)$
can be computed through 

\begin{equation}
\mathcal{S}(q)=\frac{1}{N}\sum_{i,j=1}^{N}[(\bm{\Phi}\bm{\Phi}^{\dagger})_{ii}\delta_{ij}-(|\bm{\Phi}\bm{\Phi}^{\dagger}|^{2})_{ij}]e^{\textrm{i}q(i-j)}
\end{equation}
where matrix $\bm{\Phi}$ was previously defined below Eq.$\,$\ref{eq:fid-1}.
In gapless extended phases, $\mathcal{S}(q)$ typically behaves as
$\mathcal{S}(q)=\frac{K}{2\pi}q$ at small $q$\footnote{In a Luttinger liquid phase, $K$ is the Luttinger liquid correlation
parameter. }.

\section{Results}

\subsection{Aubry-André and GPS models}

\begin{figure}[h]
\centering{}\includegraphics[width=1\columnwidth]{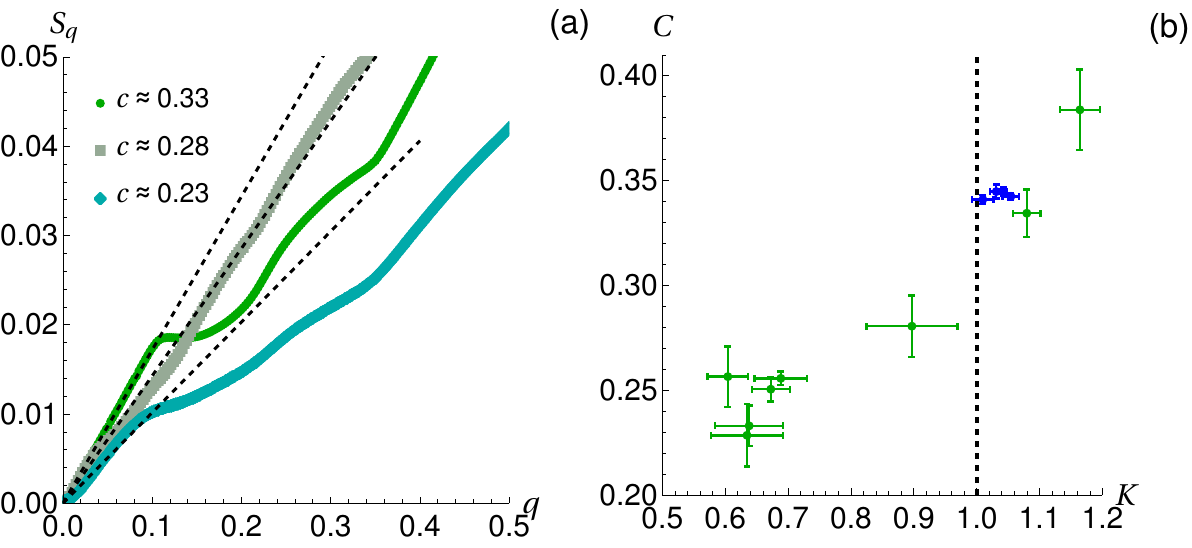}\caption{(a) Structure factor $\mathcal{S}(q)$ for the same critical points
analysed in Fig.$\,$\ref{fig:2}, with the corresponding values of
$\mathcal{C}$ indicated in the legend. The black dashed lines correspond
to the $\mathcal{S}(q)=Kq/(2\pi)$, with the $K$ computed in (b)
for the corresponding critical point. (b) $\mathcal{C}$ vs. $K$
for the different critical (green) and extended (blue) points studied
in Fig.$\,$\ref{fig:2}. $K$ was computed through $K=2\pi\mathcal{S}(\delta q)/\delta q$,
with $\delta q=20\pi/N$ and for a fixed random configuration of $\kappa$
and $\phi$ (the results of $\mathcal{S}(q)$ depend very weakly on
$\kappa$ and $\phi$). The error bars correspond to the standard
deviation of the obtained results for the system sizes $N=\{1597,2584,4181\}$.
The extended points are clustered close to $K=1$, marked with the
vertical dashed line. \label{fig:3}}
\end{figure}

We start by analysing the Aubry-André model. Some care is needed when
computing the entanglement entropy, since energy gaps are known to
open at commensurate fillings $\rho_{n}=\mod(n\tau,1),n\in\mathbb{Z}$
\citep{bernevig2013topological,PhysRevB.101.174203}. In fact, by
applying perturbation theory in the quasiperiodic potential with respect
to the homogeneous system, one finds that a gap is opened at filling
$\rho_{n}$ at order $|n|$ in perturbation theory and therefore the
gap size typically decreases with $|n|$. While in the extended phase,
gaps can only be effectively opened for $|n|\lesssim\xi$, where $\xi$
is the correlation length ($\xi=1/\log(2/V)$ in the extended phase
for the Aubry-André model \citep{AubryAndre}), at critical points/phases
$\xi$ diverges and gaps can open at any order. For a finite system,
we cannot get arbitrarily close to an incommensurate filling and we
may end up choosing a (gapped) commensurate filling for some sizes.
In order to try to avoid this problem we (i) start by choosing a number
$N'_{p}=\lfloor\rho N\rceil$ of filled states, where $\lfloor x\rceil$
rounds $x$ to the nearest integer; (ii) compute the energy gaps $\Delta_{g}^{0}=\epsilon(N'_{p})-\epsilon(N'_{p}-1)$
and $\Delta_{g}^{1}=\epsilon(N'_{p}+1)-\epsilon(N'_{p})$ , where
$\epsilon(N_{p})$ is the single-particle eigenenergy of the $N_{p}$-th
eigenstate, and choose the final number of filled states to be $N_{p}=N'_{p}-1+\arg\min_{j}\{\Delta_{g}^{j}\},\:j=0,1$.
In this way, by always choosing the smallest gap closer to the chosen
filling $\rho$ we minimize the risk of accidentally coming across
a commensurate filling for a given system. In Fig.$\,$\ref{fig:1}
we show the results for the fidelity susceptibility and entanglement
entropy obtained by carrying out the procedure just described for
the Aubry-André model, at quarter-filling ($\rho=1/4$), comparing
with the results for a close commensurate filling $\rho_{57}=\mod(57\tau,1)\approx0.228$.
In Fig.$\,$\ref{fig:1}(a) we find true criticality since $\chi_{F}$
is superextensive at the critical point. This arises from the gapless
nature of the chosen filling, as can be seen by the decreasing of
the energy gap, $\Delta_{g}$, with $N$ at the critical point {[}inset
of Fig.$\,$\ref{fig:1}(a){]}. For the commensurate filling, the
system is gapped as can be seen by the convergence of $\Delta_{g}$
in $N$, in the inset of Fig.$\,$\ref{fig:1}(b). Even though this
gap is quite small, we can see that there is clearly no divergence
in $\chi_{F}/N$ for a sufficiently large system size at the critical
point. In this case, the system is always gapped, even at the critical
point, which gives rise to an avoided criticality. These qualitative
differences naturally manifest in the scaling of the entanglement
entropy, as shown in Figs.$\,$\ref{fig:1}(c,d). While in the gapless
case, it scales as $\log N_{A}$ for $N_{A}\ll N$, only saturating
at extensively larger length scales $\Lambda\simeq N/2$ (see Eq.$\,$\ref{eq:calebrese})\textcolor{black}{,
in the gapped case, it only grows up to a $N$-independent length
scale $\xi_{g}$, inversely proportional to the gap size. In Figs.$\,$\ref{fig:1}(c,d)
we also compare the results for fillings $\rho^{(1)}=\rho_{10}$ and
$\rho^{(2)}=\rho_{57}$, where we clearly see that $S$ saturates
for $N_{A}\gtrsim\xi^{(1)}$ and $N_{A}\gtrsim\xi^{(2)}>\xi^{(1)}$
since $\Delta^{(1)}>\Delta^{(2)}$.}

From this point on we focus on incommensurate/gapless fillings. In
Fig.$\,$\ref{fig:2}(a) we compute the entanglement entropy for the
quarter-filled Aubry-André model, and for $V$ close to the critical
point. At the critical point we see that $S$ scales as $\log N_{A}$
with $\mathcal{C}\approx0.23$, slower than in the homogeneous system.
Interestingly, close to the critical point we see that $S$ closely
follows the critical behaviour for length scales $N_{A}\lesssim\xi$,
where $\xi$ is the correlation length ($\xi=1/\log(2/V)$ in the
extended phase, while $\xi=1/\log(V/2)$ in the localized phase).
For $N_{A}\gtrsim\xi$, we see a crossover to a scaling with $\mathcal{C}=1/3$
in the extended phase, as in the homogeneous case, while there is
a saturation in the localized phase. In Fig.$\,$\ref{fig:2}(b),
we compute the scaling of $S$ in different critical points/phases,
comparing with the results obtained deep in the extended phase. We
see that the scaling of $S$ in the critical phase can vary significantly,
and even closely resemble the homogeneous/extended scaling. This is
in contrast with previous results obtained for the Aubry-André model
with long-range hoppings for which $\mathcal{C}$ was always found
to be significantly lower than in the homogeneous or extended case,
when the fillings were chosen in regions of multifractal eigenstates.
We also note that even though $S$ grows with $\log N_{A}$, it also
shows small oscillations. These oscillations are expected: in fact
log-periodic oscillations were previously found at the critical point
of the half-filled Aubry-André model with $\tau=1/\sqrt{2}$ \citep{goncalves2023shortrange}
and also in aperiodic spin chains \citep{Juhasz_2007,Igloi_2007}.
In Figs.$\,$\ref{fig:2}(c,d), we compute $\mathcal{C}$ at localization-delocalization
transitions (c) and inside the critical phase (d) by fitting $S(N_{A})$
to Eq.$\,$\ref{eq:calebrese}. We find that $\mathcal{C}$ can take
a wide range of non-universal values. Interestingly, for localization-delocalization
transitions at filling $\rho=1/4$ we find that the scaling of the
entanglement entropy for the Aubry-André and GPS models is compatible,
in agreement with the critical point universality proposed in Ref.$\,$\citep{goncalves2023shortrange}.
This again suggests that at extended-localized transitions the behaviour
of entanglement entropy is universal for significantly different models
at fixed fillings, as conjectured in Ref.$\,$\citep{goncalves2023shortrange}. 

We now try to establish a connection between the scaling of $S$ and
other physical properties. At critical points, correlation functions
of the type $\langle\mathcal{O}_{i}\mathcal{O}_{j}\rangle$ are expected
to decay in a power-law fashion as $\langle\mathcal{O}_{i}\mathcal{O}_{i+\epsilon}\rangle\sim|\epsilon|^{-y_{\mathcal{O}}}$
for large enough $|\epsilon|$. Taking the simplest correlation function,
$\langle c_{i}^{\dagger}c_{i+\epsilon}\rangle$, we find that $\langle c_{i}^{\dagger}c_{i+\epsilon}\rangle\sim|\epsilon|^{-1}$
in the extended phase with the interesting possible formation of moiré
patterns (see Appendix \ref{sec:AppendixB}); while in the localized
phase $\langle c_{i}^{\dagger}c_{i+\epsilon}\rangle\sim e^{-|\epsilon|/\xi}$,
where $\xi$ is the localization length. At critical phases/points,
however, the are very large fluctuations as function of $\epsilon$
and it becomes challenging to define the power-law exponent. In fact,
the scaling of the maxima of these fluctuations is compatible with
$|\epsilon|^{-1}$. Therefore, no significant distinctions in the
scalings of $g_{r}(\epsilon)$ were found in the critical regions.
The connection between $S$ and the general behaviour of correlation
functions is therefore more subtle. In fact, a clear relation between
the entanglement entropy and the low-momentum scaling of the structure
factor $\mathcal{S}(q)$, given in Eq.$\,$\ref{eq:Sq}, can be found.
In Fig.$\,$\ref{fig:3}(a) we can see that only the slope of the
scaling, $K$ (see below Eq.$\,$\ref{eq:Sq}), changes at critical
points, while the linear scaling in $q$ still holds as in the extended
phase, up to small oscillations. By computing $K$ for the different
critical and extended points that we studied in Fig.$\,$\ref{fig:2},
and comparing the results with $\mathcal{C}$, we verified that $K$
increases with $\mathcal{C}$ as shown in Fig.$\,$\ref{fig:3}(b).
While at extended points we have $K\approx1$ (blue points in Fig.$\,$\ref{fig:3})
as in the homogeneous case, in the critical case the values of $K$
can vary significantly. 

\subsection{Quasiperiodic Kitaev chain}

\begin{figure}[h]
\centering{}\includegraphics[width=1\columnwidth]{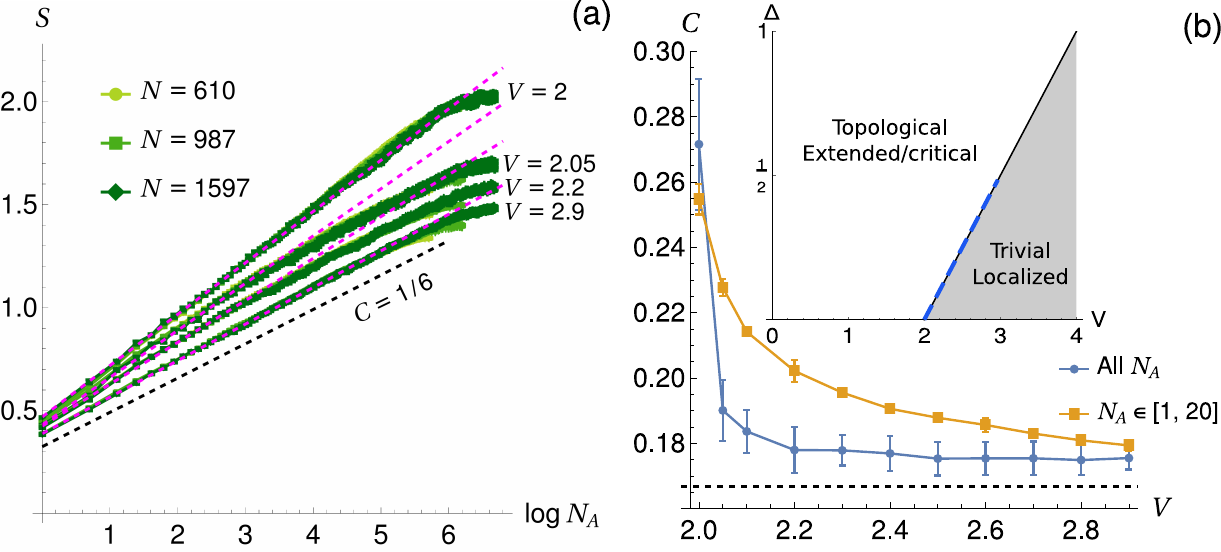}\caption{(a) $S$ for the quasiperiodic Kitaev chain, for $\mu=0$ and different
$(V,\Delta)$ across the topological transition shown in the inset
of (b), given by $\Delta(V)=V/2-1$. The values of $V$ are indicated
close to the corresponding curves. The dashed magenta lines correspond
to linear fits made for $N_{A}\protect\leq20$. (b) $\mathcal{C}$
obtained by fitting $S$ to Eq.$\,$\ref{eq:calebrese} for all $N_{A}$
(blue) and by making a linear fit of the $(\log N_{A},S)$ data only
for $N_{A}\protect\leq20$ (yellow). The data points were computed
by averaging the results obtained for the system sizes $N=\{610,987,1597\}$
and the error bars correspond to the standard deviation. The inset
contains the topological phase diagram in the $(V,\Delta)$ plane.
The blue dashed line in the inset indicates the explored range of
the critical line. \label{fig:4}}
\end{figure}

\begin{figure}[h]
\begin{centering}
\includegraphics[width=1\columnwidth]{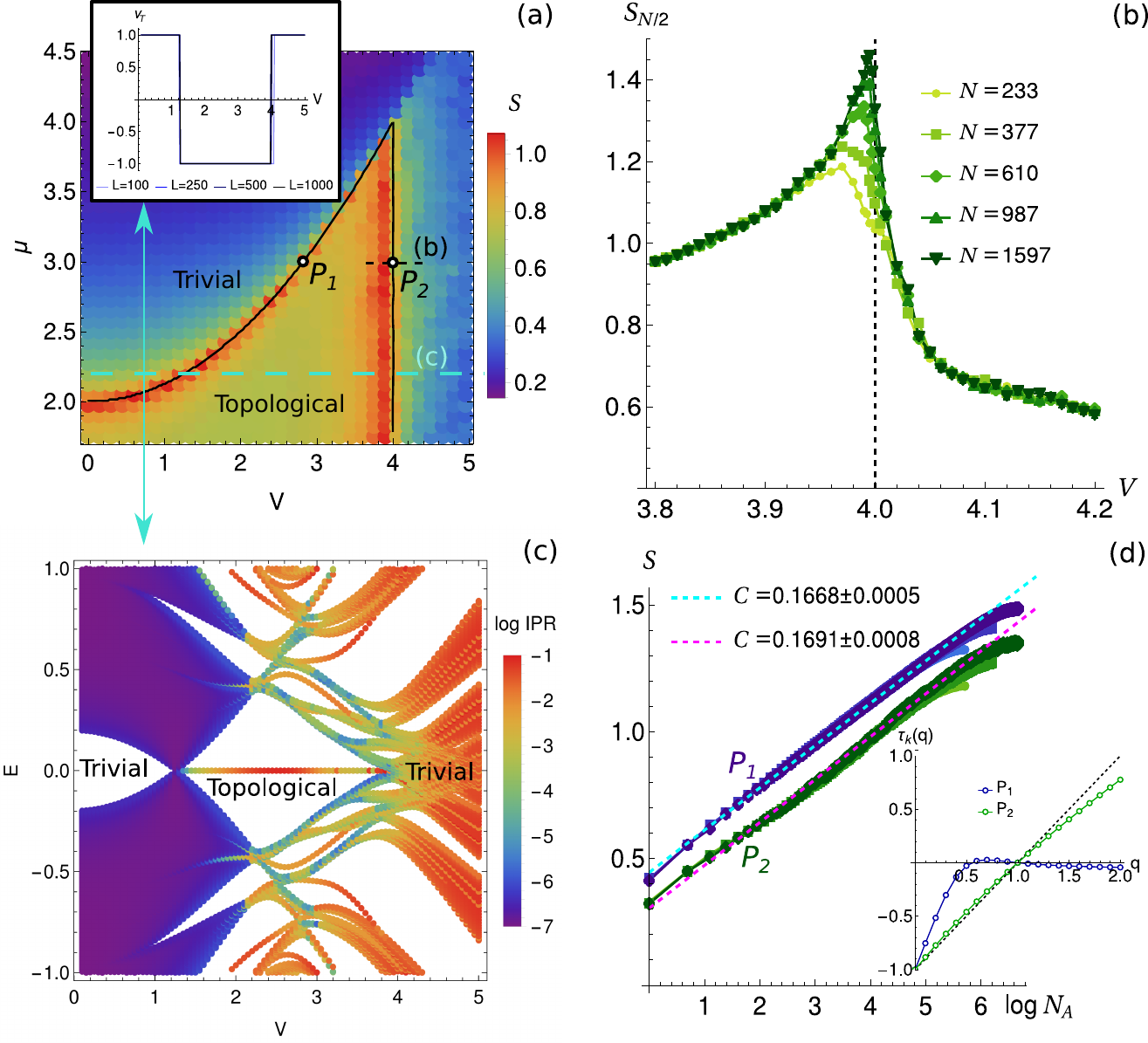}
\par\end{centering}
\caption{Results for $\Delta=1$. (a) $S$ for $N=233$ and $N_{A}=[N/6]$,
with each point corresponding to an average over 75 realizations.
The full black line shows the phase boundaries obtained through the
topological invariant in Eq.$\,$\ref{eq:topo_invariant-1}. The inset
show the results for this topological invariant at a fixed $\mu$
cut $(\mu=2.2)$ marked by the cyan dashed line. (b) Finite-size scaling
of $S$ for variable $N$ and $N_{A}=[N/2]$. We averaged over $N_{c}=1800$
configurations for $N\protect\leq987$ and $N_{c}=252$ for $N=1597$.
(c) IPR results for $N=987$, with open boundary conditions, and for
parameters at the cyan dashed cut in (a). (d) $S$ at critical points
$P_{1}$ and $P_{2}$ indicated in (a). The lighter, intermediate
and darker colors correspond respectively to $N=610,987,1597$, with
the results respectively averaged over $N_{c}=1800,1800,252$ configurations.
The inset contains the results for the multifractal exponent $\tau_{k}(q)$,
computed by fitting $\textrm{IPR}_{k}(q)$ for $N\in[144,2584]$.
We used $N_{c}\in[25-725]$ configurations of $\phi$ for $P_{2}$
and a single random configuration for $P_{1}$ (the $\phi$-dependence
is neglegible in this case). We note that for $P_{1}$, $\tau_{k}(q)$
deviates significantly from $0$ for small $q$, which we attribute
to finite-size effects.\label{fig:5}}
\end{figure}

The quasiperiodic Kitaev chain has a well-known phase diagram in the
$\Delta-V$ plane for $\mu=0$ \citep{PhysRevLett.110.146404}, shown
in the inset of Fig.$\,$\ref{fig:4} (see also Ref.$\,$\citep{PhysRevB.93.104504}
for full localization phase diagram and Ref.$\,$\citep{PhysRevB.108.L100201}
for exact analytical solution). Here we computed the entanglement
entropy at the critical line of the transition between a topological
phase with critical eigenstates, having majorana zero modes, and a
trivial localized phase. The results for $\mathcal{C}$ are shown
in Fig.$\,$\ref{fig:4}. For $(V,\Delta)=(2,0)$, $\mathcal{C}$
has the value of the Aubry-André critical point ($\mathcal{C}\approx0.26$
at half-filling \citep{PhysRevB.102.064204}). However, for finite
$\Delta$, $\mathcal{C}$ quickly becomes smaller, getting close to
the homogeneous Kitaev chain's value, $\mathcal{C}=1/6$, with the
most significant deviations arising only at small $N_{A}$. This can
be observed by fitting the scaling of the entanglement entropy at
small $N_{A}$ - Fig.$\,$\ref{fig:4}(b) - for which we see a significant
$V$-dependence. When fitting $S$ to Eq.$\,$\ref{eq:calebrese}
using all $N_{A}$, we found that $\mathcal{C}$ is essentially constant
for different critical points of the transition, in contrast to the
critical points of the unpaired models, for which $\mathcal{C}$ varied
significantly.

In order to understand whether the almost constant value of $\mathcal{C}$
was a special feature of the analysed transition, we also studied
the $\mu\neq0$ case. In this case we can have reentrant topological
transitions, similar to what is observed for topological Anderson
insulators, in the disordered case \citep{PhysRevLett.102.136806,universe5010033}.
This is shown in Fig.$\,$\ref{fig:5}(a) for $\Delta=1$. In this
figure, we can see that we can start in a trivial phase, at $\mu>2$,
and transition into a topological phase by increasing $V$. The topological
phase contains zero-energy modes as expected, which we illustrate
in Fig.$\,$\ref{fig:5}(c). At an even larger value of $V$, we have
a new topological transition back into a trivial phase. These two
topological transitions are however of different nature in terms of
localization properties. Exactly at the transition point the eigenstates
are ballistic in the first and multifractal in the second. This is
shown in the inset of Fig.$\,$\ref{fig:5}(d), where we plotted $\tau_{k}(q)$
for examples of these transitions, at points $P_{1}$ and $P_{2}$,
indicated in Fig.$\,$\ref{fig:5}(a). We can see, however, that the
scaling of the entanglement entropy is similar for both transitions,
apart from the existence of oscillations in the critical case, as
in the other studied critical points. 

\section{Discussion}

In this work, we characterized in detail the entanglement entropy
in different models of one-dimensional fermionic quasiperiodic systems,
focusing on multifractal critical points/phases. In the absence of
pairing, we found that the entanglement entropy follows the expected
behaviour for one-dimensional critical systems, but can show possible
small oscillations. Similar log-periodic oscillations were previously
found in Ref.$\,$\citep{Juhasz_2007,Igloi_2007} for aperiodic spin
chains and in Ref.$\,$\citep{goncalves2023shortrange} both for the
non-interacting and interacting Aubry-André model. In the absence
of pairing, we found that the characteristic coefficient $\mathcal{C}$
that governs the scaling of the entanglement entropy with subsystem
size $N_{A}$ and total size $N$ depends significantly on the model
parameters and electron filling. At some critical points, we observed
that the entanglement entropy even behaves similarly to the homogeneous
case, with $\mathcal{C}\approx1/3$. This is in contrast with previous
studies where $\mathcal{C}$ was always found to be significantly
smaller at critical multifractal phases/points, compared to the homogeneous
case \citep{PhysRevB.103.075124,PhysRevB.102.064204}. Although there
are significant variations of $\mathcal{C}$, we find that compatible
values are obtained for the Aubry-André and GPS models at critical
points of localization-delocalization transitions, for a fixed filling.
This is in agreement with Ref.$\,$\citep{goncalves2023shortrange},
where a universal behaviour of the entanglement entropy was found
at the localization transitions of different half-filled quasiperiodic
chains. We also compared the behaviour of the entanglement entropy
with correlation functions and found that $\mathcal{C}$ was closely
related with the long-wavelength behaviour of the momentum structure
factor. 

For critical quasiperiodic Kitaev chains, we found that the addition
of even small pairing terms is highly relevant for the behaviour of
the entanglement entropy. Independently of studied critical points,
we find that it behaves similarly to the homogeneous critical Kitaev
chain, where $\mathcal{C}=1/6$, departing from the non-universal
larger values of $\mathcal{C}$ computed in the unpaired case. The
most significant deviations from $\mathcal{C}=1/6$ only occur at
small length scales, when the critical point contains multifractal
eigenstates.

Future interesting studies include to address the impact of interactions
on the entanglement entropy, in critical phases with multifractal
eigenstates. In Ref.$\,$\citep{goncalves2023shortrange}, short-range
spinless interactions were found to be irrelevant at critical points
between extended and localized phases, so we expect that our results
at these transitions also hold in the interacting case. However, interactions
can be relevant in multifractal critical phases of quasiperiodic systems
\citep{goncalves2023incommensurability}, which may change the behaviour
of the entanglement entropy. 

\bibliographystyle{apsrev4-1}
\bibliography{1D_Hidden_SD_Paper}




\clearpage{}

\appendix

\section{Quasiperiodic Kitaev chain: clean limit and topological properties}

\begin{figure}[h]
\begin{centering}
\includegraphics[width=1\columnwidth]{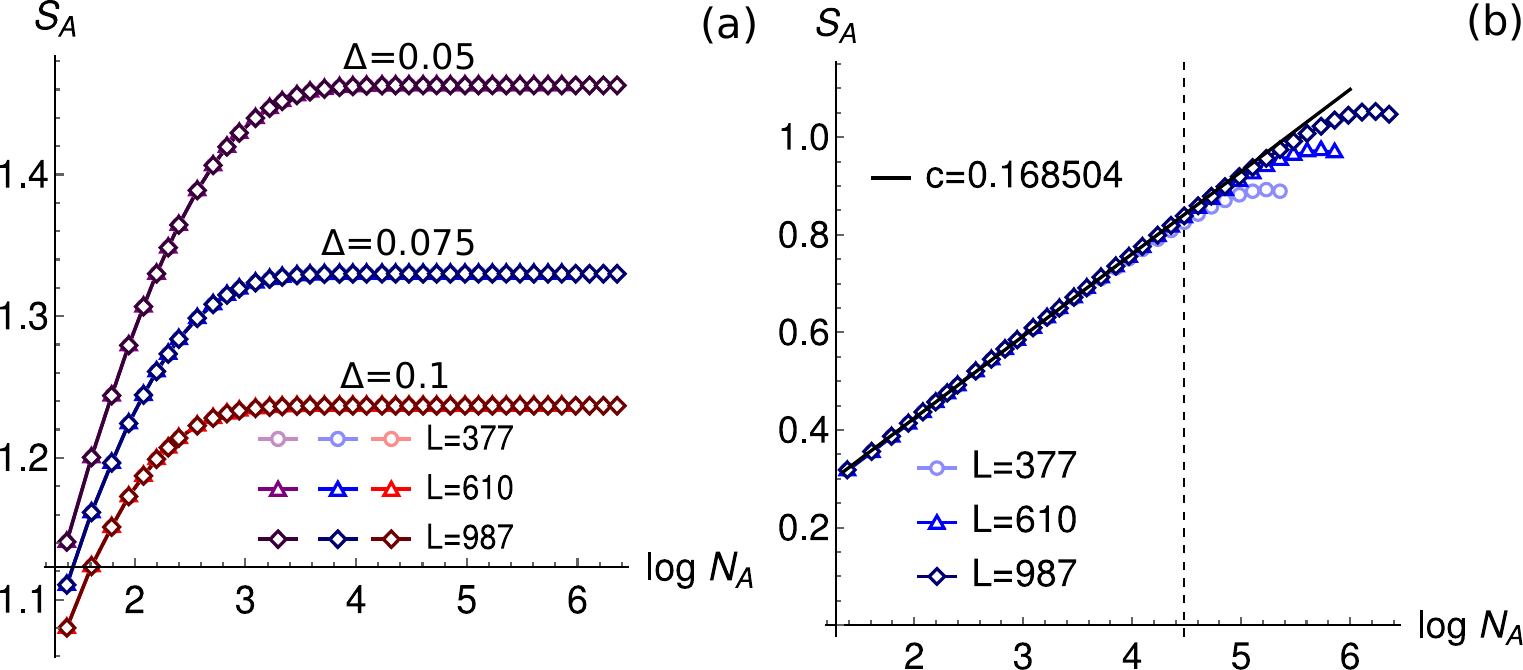}\caption{(a) Entanglement entropy for the homogeneous Kitaev chain, for $\mu=0$
and variable $\Delta$. Note that the energy gap increases with $\Delta$.
Therefore, for larger $\Delta$, $S$ saturates for smaller $N_{A}$,
of the order of the inverse gap. (b) Entanglement entropy for the
clean Kitaev chain, for $\Delta=0.1,\mu=2$ (critical point). $S_{A}$
scales with the correct pre-factor $\mathcal{C}\approx1/6$ consistent
with the central charge $c=1/2$ of the Kitaev model.\label{fig:kitaev_first}}
\par\end{centering}
\end{figure}

In this section, we provide additional details on how the topological
properties of the quasiperiodic Kitaev chain were studied in the main
text. For $V=0$, this model is homogeneous and exhibits a phase transition
at $|\mu|=|t|$, for any $\Delta$ \citep{DeGottardi_2011}. At this
point, the entanglement entropy scales with $\mathcal{C}=1/6$ \citep{PhysRevLett.90.227902}.
Otherwise, the system is gapped and the entanglement entropy saturates
for large enough $N_{A}$. We reproduce these results, in Fig.$\,$\ref{fig:kitaev_first}.

In order to study the topological properties of the model, in particular
for the phase diagram in Fig.$\,$\ref{fig:5}(a), we used a topological
invariant first defined in Ref.$\,$\citep{DeGottardi_2011}. We now
provide some additional details on this quantity. Let us introduce
the Majorana fermions $a_{n}$ and $b_{n}$ such that $c_{n}=(a_{n}+ib_{n})/2$.
$a_{n}$ and $b_{n}$ satisfy $\{a_{n},b_{n}\}=0$ and $\{a_{n},a_{m}\}=\{b_{n},b_{m}\}=2\delta_{mn}$
(and therefore $a_{n}^{2}=b_{n}^{2}=1$). The Hamiltonian for the
quasiperiodic Kitaev chain becomes:

\begin{equation}
\begin{aligned}H=\frac{\textrm{i}}{2}\sum_{n} & \Big((-t+\Delta)a_{n}b_{n+1}+(t+\Delta)b_{n}a_{n+1}\\
 & +[V\cos(2\pi\tau n+\phi)-\mu]a_{n}b_{n}\Big)+{\rm cte}
\end{aligned}
\label{eq:Ham_maj}
\end{equation}

The Majorana zero modes are represented by the operators $\Gamma_{a}=\sum_{n}\alpha_{n}a_{n}$
and $\Gamma_{b}=\sum_{n}\beta_{n}b_{n}$, whose amplitudes $\alpha_{n}$
and $\beta_{n}$ satisfy the zero-energy equations of motion for the
Hamiltonian in Eq.$\,$\ref{eq:Ham_maj}:  

\begin{equation}
\begin{aligned}(t-\Delta)\alpha_{n-1}+(t+\Delta)\alpha_{n+1}-V_{n}\alpha_{n} & =0\\
-(t+\Delta)\beta_{n-1}-(t-\Delta)\beta_{n+1}+V_{n}\beta_{n} & =0
\end{aligned}
\end{equation}

\noindent where $V_{n}=V\cos(2\pi\tau n+\phi)-\mu$. We can write
these decoupled equations in the transfer matrix form:

\begin{equation}
\left(\begin{array}{c}
\alpha_{n+1}\\
\alpha_{n}
\end{array}\right)=A_{n}\left(\begin{array}{c}
\alpha_{n}\\
\alpha_{n-1}
\end{array}\right),\textrm{ }A_{n}=\left(\begin{array}{cc}
\frac{V_{n}}{\Delta+t} & \frac{\Delta-t}{\Delta+t}\\
1 & 0
\end{array}\right)
\end{equation}

\begin{equation}
\left(\begin{array}{c}
\beta_{n+1}\\
\beta_{n}
\end{array}\right)=B_{n}\left(\begin{array}{c}
\beta_{n}\\
\beta_{n-1}
\end{array}\right),\textrm{ }B_{n}=\sigma_{x}A_{n}^{-1}\sigma_{x}
\end{equation}

\noindent where $\sigma_{x}$ is a Pauli matrix. Considering a semi-infinite
chain starting at site $n=1$, we have $\alpha_{0}=\beta_{0}=0$.
At site $n=N+1$ we have 

\begin{equation}
\left(\begin{array}{c}
\alpha_{N+1}\\
\alpha_{N}
\end{array}\right)=\mathcal{A}_{L}\left(\begin{array}{c}
\alpha_{1}\\
0
\end{array}\right),\textrm{ \ensuremath{\mathcal{A}_{N}}}=\prod_{n=1}^{N}A_{n}
\end{equation}

In the same way, we can define

\begin{equation}
\textrm{ \ensuremath{\mathcal{B}_{N}}}=\prod_{n=1}^{N}B_{n}
\end{equation}

In order to have Majorana modes localized at the edges of the chain,
both the eigenvalues of $\mathcal{A}_{N}$ should be either smaller
or larger than one in magnitude as $N\rightarrow\infty$. In the former
case, we have a localized $a$-mode, while in the latter, we have
a localized $b$-mode (at the left boundary). If only one of the eigenvalues
of $\mathcal{A}_{N}$ is larger than unity, then both the $a$- and
$b$-modes are not normalizable and we cannot have localized Majorana
modes.

Based on these considerations we can define a topological invariant
as in Ref.$\,$\citep{DeGottardi_2011}:

\begin{equation}
\nu_{T}=-(-1)^{n_{f}}\label{eq:topo_invariant}
\end{equation}

\noindent where $n_{f}$ is the number of eigenvalues with magnitude
smaller than unity. For $n_{f}=0,2$, $\nu_{T}=-1$ and the phase
is topological. For $n_{f}=1$, $\nu_{T}=1$ and the phase is trivial.

\section{Single-particle correlation matrix in the Aubry-André model}

\label{sec:AppendixB}

We finally provide additional results on single-particle correlation
functions for the Aubry-André model. We define the real and momentum-space
single-particle correlation matrices as 

\begin{equation}
\chi_{rr'}=\langle c_{r}^{\dagger}c_{r'}\rangle
\end{equation}

\begin{equation}
\Lambda_{\kappa\kappa'}=\frac{1}{N}\sum_{rr'}e^{\textrm{i}(\kappa r-\kappa'r')}\chi_{rr'}
\end{equation}

For translational invariant systems, $\Lambda_{\kappa\kappa'}$ is
diagonal. In fact, we have $n_{\kappa}\equiv\Lambda_{\kappa\kappa}=\Theta[-(\kappa-\kappa_{F})]-\Theta[-(\kappa+\kappa_{F})]$,
where $\kappa_{F}$ is the Fermi momentum. An interesting question
is what happens to $\Lambda_{\kappa\kappa'}$ when we switch on the
quasiperiodic perturbation. On the one hand, gaps will open and $n_{\kappa}$
will stop being the Heaviside function. On the other, off-diagonal
elements appear in $\Lambda_{\kappa\kappa'}$ because translational
invariance is broken. 

We take the Aubry-André model as an example. For the Aubry-André model,
different momenta are coupled if $\kappa-\kappa'=2\pi\tau j,\textrm{ }j\in\mathbb{Z}$,
or, for a commensurate approximant with $\tau_{n}=F_{n-1}/F_{n}$,
if $\kappa-\kappa'=2\pi\tau_{n}j,\textrm{ }j=0,\cdots,N-1$ \citep{AubryAndre}.
Therefore, it is convenient to use $\kappa_{j}=2\pi\tau_{c}j$ and
$\kappa_{j}'=2\pi\tau_{c}j'$, with $j,j'=0,\cdots,N-1$ and define 

\begin{equation}
\Lambda_{jj'}=\frac{1}{N}\sum_{rr'}e^{i2\pi\tau_{n}(jr-j'r')}\chi_{rr'}
\end{equation}

Note that due to periodic boundary conditions, $\Lambda_{j+mN,j'+lN}=\Lambda_{jj'}$
for $m,l\in\mathbb{Z}$. In order to have an idea on how $\Lambda_{jj'}$
behaves for $j\neq j'$, we define

\begin{equation}
g_{\kappa}(\epsilon)=\frac{1}{N}\sum_{j}\Lambda_{j,j+\epsilon}
\end{equation}

For $\epsilon=0$, we have that $g(0)=\Tr(|\bm{\Lambda}|)=\nu$, where
$\nu$ is the filling. We define an analogous quantity for the real-space
correlation function:

\begin{equation}
g_{r}(\epsilon)=\frac{1}{N}\sum_{r}\chi_{r,r+\epsilon}
\end{equation}

\begin{figure*}[t]
\centering{}\includegraphics[width=0.8\paperwidth]{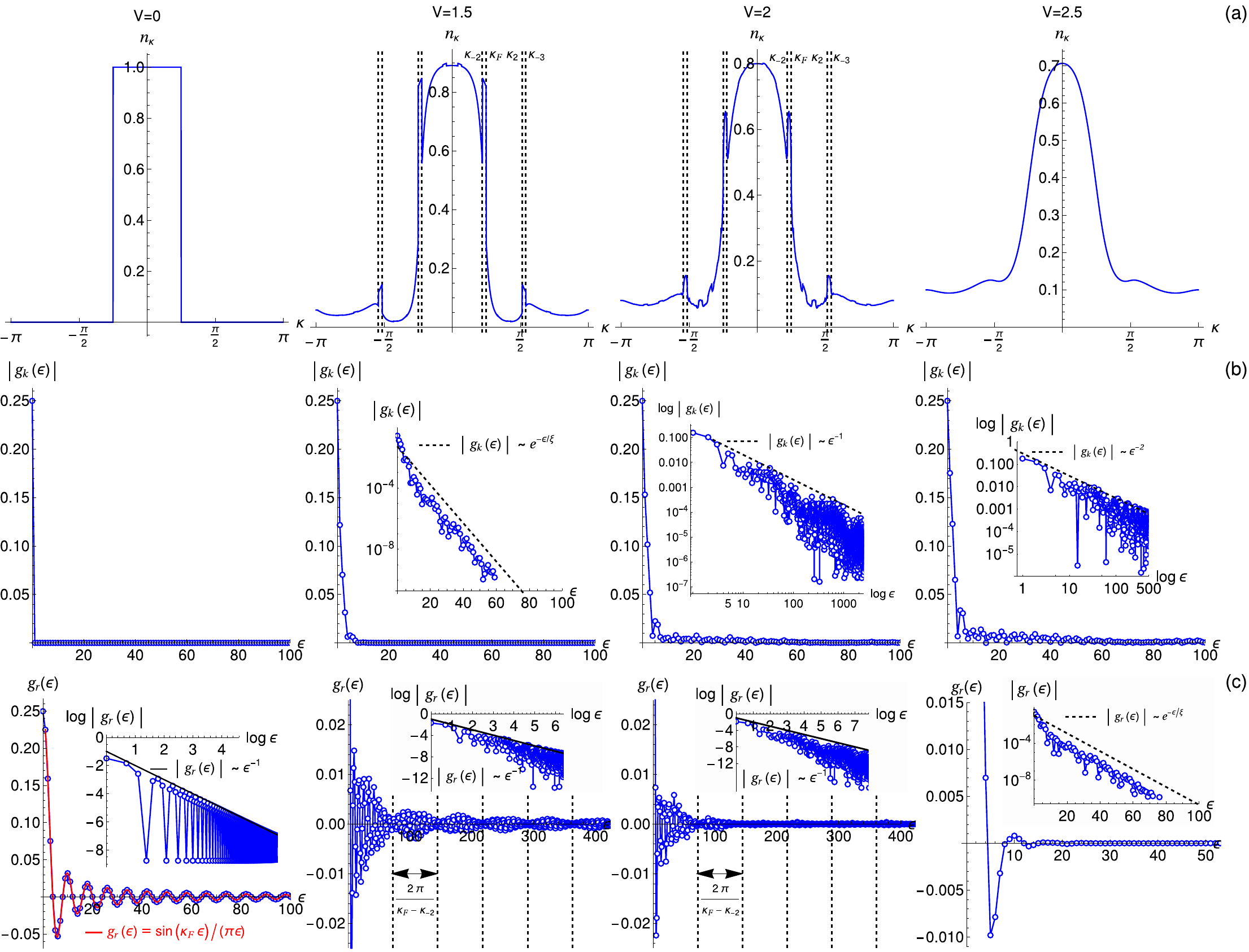}\caption{Results for $n_{\kappa}$ (a), $g_{\kappa}(\epsilon)$ (b) and $g_{r}(\epsilon)$
(c), for the quarter-filled Aubry-André model, for different values
of $V$ in each column (indicated above the corresponding column).
In (a) we label some values $\kappa_{n}=\kappa_{F}+2\pi\tau n,n\in\mathbb{Z}$.
(b) The first inset from left to right is a log plot, showing that
$g_{\kappa}(\epsilon)$ decays exponentially with $\epsilon$ for
$V=1.5$, with a exact correlation length $\xi=1/\log(2/V)$, while
the last two insets are log-log plots showing that $|g_{\kappa}(\epsilon)|$
decays as a power-law envelope for $V=2$ and $V=2.5$. (c) The red
line in the leftmost figure corresponds to the exact result $g_{r}(\epsilon)=\sin(\kappa_{F}\epsilon)/(\pi\epsilon)$,
for $V=0$. The first three insets correspond to log-log plots showing
the power-law envelop decay of $|g_{r}(\epsilon)|$, while the last
plot is a log plot showing the exponential decay for $V=2.5$ with
the exact localization length $\xi=1/\log(V/2)$. We note that there
are moiré patterns that can be noted in the middle figures. These
have a periodicity $\Delta\epsilon=2\pi/(\kappa_{F}-\kappa_{-2})$,
where $\kappa_{F}$ is the Fermi momentum and $\kappa_{-2}$ is indicated
in (a). All the dashed/full lines in the insets are guides to the
eye. The results are all for $N=1597,\phi=1.123,\kappa=0.001$, except
for $V=2$ where we used $N=4181$. A very small $\kappa$ was used
simply to break fermi-level degeneracies. Note that since this introduces
a small imaginary part in $\chi_{rr'}$, we just took the real part
of $g_{r}(\epsilon)$ in the figures. \label{fig:corr_mat_results}}
 
\end{figure*}

We provide a summary of the results in Fig.$\,$\ref{fig:corr_mat_results},
for the quarter-filled Aubry-André model. For $V=0$, states with
definite crystal momentum $\kappa$ form the eigenbasis of the Hamiltonian
and therefore $\Lambda_{\kappa\kappa'}$ is a diagonal matrix {[}which
can be seen from the results for $g_{\kappa}(\epsilon)${]}, with
$n_{\kappa}\equiv\Lambda_{\kappa\kappa}=\begin{cases}
1, & \textrm{ }|\kappa|\leq\kappa_{F}\\
0, & \textrm{ otherwise}
\end{cases}$. From this, one can easily obtain that $g_{r}(\epsilon)=\sin(\kappa_{F}\epsilon)/(\pi\epsilon),\textrm{ }\epsilon>0$.
For $V>0$ within the extended phase, we can see that $\Lambda_{\kappa\kappa'}$
is no longer diagonal, but $g_{\kappa}(\epsilon)$ decays exponentially
away from the diagonal with a correlation length $\xi=1/\log(2/V)$,
the exactly known correlation length for the extended phase of the
Aubry-André model. Furthermore, $n_{\kappa}$ develops additional
discontinuities at wave vectors $\kappa_{n}=\kappa_{F}+2\pi\tau n,n\in\mathbb{Z}$,
to which the Fermi momentum $\mathsf{k}_{F}$ couples in perturbation
theory \citep{bernevig2013topological}. We note however that in practice,
the coupling does not occur for any $n$ due to the finite correlation
length $\xi$: no gaps open for $|n|\gg\xi$. This can give rise to
curious situations as the one that we illustrate in Fig.$\,$\ref{fig:corr_mat_results}
for the quarter-filled Aubry-André model. In this case there is a
small-order $\kappa_{n}$ corresponding to $n=-2$ very close to the
Fermi momentum $\kappa_{F}$ \footnote{We note that even though $\kappa_{F}$ is only well-defined for $V=0$,
$n_{\kappa}$ still shows a sharp decrease at $\kappa=\kappa_{F}$
in the extended (and even critical) phase of the Aubry-André model.}, which gives rise to the small momentum windows with large $n_{\kappa}$
shown in Fig.$\,$\ref{fig:corr_mat_results}(a). This translates
in the formation of a real-space beating (moiré) pattern with wave
vector $\Delta\kappa=\kappa_{F}-\kappa_{-2}$. This can be simply
understood if one just considers the diagonal contributions of $\Lambda_{\kappa\kappa'}$,
given by $\chi_{rr'}\approx N^{-1}\sum_{\kappa}e^{-\textrm{i}\kappa(r-r')}n_{\kappa}$.
Taking the large contributions at $\kappa\in[\kappa_{-2},\kappa_{F}]\bigcup[-\kappa_{F},-\kappa_{-2}]$,
that we call $\chi_{rr'}^{(\Delta\kappa)}$, and assuming that they
are approximately a constant $\tilde{n}_{\Delta\kappa}$, we get $\chi_{rr'}^{(\Delta\kappa)}\approx(2\pi)^{-1}\tilde{n}_{\Delta\kappa}\big(\int_{-\kappa_{F}}^{-\kappa_{-2}}+\int_{\kappa_{-2}}^{\kappa_{F}}\big)d\kappa e^{-\textrm{i}\kappa(r-r')}\sim\cos[\kappa_{F}(r-r')]+\cos[\kappa_{-2}(r-r')]$,
from which the beating effect seen in $g_{r}(\epsilon)$, in Fig.$\,$\ref{fig:corr_mat_results}(c),
is immediately captured. At the critical point $V=2$, the correlation
length diverges and the off-diagonal elements of $\Lambda_{\kappa\kappa'}$
start to decay in a power-law fashion, as can be seen from the results
of $|g_{\kappa}(\epsilon)|$. In this case, the beating effect in
$g_{r}(\epsilon)$ just mentioned becomes less obvious due to the
more complex structure of the momentum-space correlation matrix. Interestingly,
using the Aubry-André duality, one can easily check that the real-
and momentum-space matrices $\chi_{rr'}$ and $\Lambda_{\kappa\kappa'}$
become equal at the critical point. Finally, in the localized phase,
the off-diagonal elements of $\chi_{rr'}$ decay exponentially with
the localization length $\xi=1/\log(V/2)$, while the off-diagonal
elements of $\Lambda_{\kappa\kappa'}$ decay with a power-law envelope. 

\begin{figure*}[h]
\centering{}\includegraphics[width=0.8\textwidth]{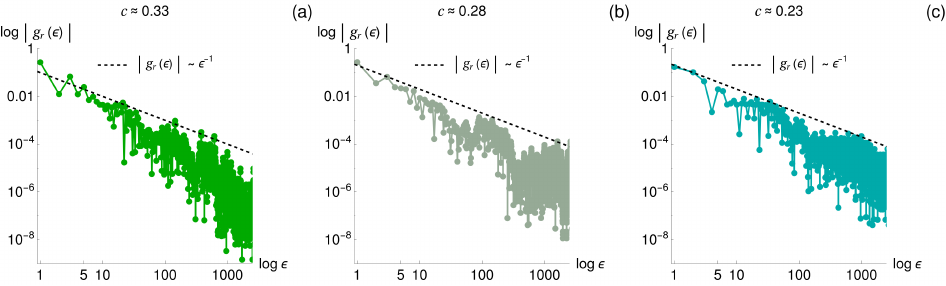}\caption{$|g_{r}(\epsilon)|$ at the different critical points also considered
in Fig.$\,$\ref{fig:2}(b). In particular: (a) $V=1,\alpha=2,\rho=0.47$;
(b) $V=1,\alpha=2,\rho=0.538$; (c) $V=1,\alpha=0.539,\rho=0.25$.
The computed entanglement entropy coefficient $\mathcal{C}$ is indicated
above the figure of the corresponding critical point. The dashed black
lines are guides to the eye corresponding to a $|g_{r}(\epsilon)|\sim\epsilon^{-1}$
scaling. \label{fig:g_epsilon_critical_points}}
\end{figure*}

From Fig.$\,$\ref{fig:corr_mat_results}, we can see that the behaviour
of the real- and momentum-space correlation matrices at the critical
point is highly non-trivial. In fact, we show in Fig.$\,$\ref{fig:g_epsilon_critical_points}
that there is no clear distinction in the scaling of $g_{r}(\epsilon)$
for example critical points that have a significantly different scaling
of the entanglement entropy $\mathcal{C}$.

\end{document}